\documentclass[a4paper,10pt]{article}

\usepackage[a4paper, total={6.5in, 8.5in}]{geometry}
\usepackage[utf8]{inputenc}
 \usepackage{graphics}
 \usepackage{graphicx}
 \usepackage{setspace}

\usepackage{amssymb}
\usepackage{amsmath}
\usepackage{amsthm}

\title{Effects of the pseudo-gap on the field-induced kinetic energy density of 
Bi$_2$Sr$_2$CaCu$_2$O$_{8+\delta}$ single crystals}
\author{L. F. Lopes$^1$, J. P. Pe\~na$^1$, M. A. Tumelero$^1$, J. Schaf$^1$, V. N. Vieira$^2$,  P. Pureur$^1$}

\begin{document}

\maketitle
\begin{center}
 
$^1$Instituto de F\'isica, Universidade Federal do Rio Grande do Sul, ZIP 91501-970 Porto Alegre, RS, Brazil\\
$^2$Instituto de F\'isica e Matem\'atica, Universidade Federal de Pelotas, ZIP 96010-900 Pelotas, RS, Brazil
\end{center}

\doublespace

\begin{abstract}

We report on magnetization experiments from which we obtain the field-induced kinetic energy density, $E_k$,
in the superconducting phase of several Bi$_2$Sr$_2$CaCu$_2$O$_{8+\delta}$  single crystal samples. 
The kinetic energy magnitude changes according to the characteristic reduction of the single-electron density 
of states produced by the pseudogap in the underdoped limit. Moreover, a remarkable peak of $E_k$ occurring at the specific holes density $p\sim0.18$
is related to a van Hove singularity  due to the pseudogap closure. 
We also extracted the superfluid density, $\rho_s$.
We conclude that  $E_k$ and $\rho_s$ are related to the pseudogap energy scale.
This result is understood as an evidence of the coexistence between superconductivity and the pseudogap phenomenon in 
the Bi$_2$Sr$_2$CaCu$_2$O$_{8+\delta}$ cuprate compound. 


\end{abstract}

\section{Introduction}

The pseudogap phenomenon has been a longstanding subject in the physics of the high temperature superconducting cuprates (HTSC). 
This property is characterized by a sharp depression of the single-electron density of states (DOS) at the Fermi level and manifests
more vigorously in the underdoped specimens. Moreover, the pseudogap problem has been shown to be closely related  to the interplay between 
superconductivity and spin or charge degrees of freedom. Experimentally, the pseudogap effects arise below a characteristic temperature $T^*$,
also known as the pseudogap temperature. 
For the most severely underdoped samples $T^*$  is sensibly higher than the superconducting critical temperature $T_c$; 
thus, the experimental manifestations of the pseudogap in photoelectron spectroscopy  ~\cite{kordyuk, vishik}, specific 
heat  ~\cite{loram-mirza, loram-luo}, nuclear magnetic resonance (NMR) ~\cite{warren}, 
tunneling conductance  ~\cite{renner}, 
transport ~\cite{ding} and optical properties ~\cite{tajima}  of the HTSC, are
commonly observed in their normal phase (see Ref. ~\cite{hashimoto} for an extended review on the pseudogap).

Although the presence of the pseudogap has been widely reported in several HTSC, a complete physical description of this phenomenon is still missing.
Experimentally, the fact that the region where the pseudogap is observed extends mainly into the 
underdoped part of the temperature vs. holes concentration phase diagram has been reported as an universal characteristic of the HTSC ~\cite{tallon-bernhard}.
It is not clear, nonetheless, whether the $T^*(p)$ curve represents a true 
phase transition boundary delimiting a pseudogap phase, or if it only identifies a crossover line 
occurring in a more or less broad temperature interval. In the underdoped part of the $T$ vs. $p$ diagram, 
where $T^*>T_c$, the doping dependence of $T^*$ is fairly well known ~\cite{tallon-loram, naquib}.
On the other hand, the evolution of the  $T$ vs. $p$ line across the optimal and overdoped regions is a matter of controversy.

In the last 30 year’s numerous theories to describe the pseudogap have been proposed, most of them converging to one of three different scenarios. 
In the first one, also known as phase-fluctuation scenario (or preformed pairs scenario), the pseudogap is treated as a consequence of preformed
incoherent electron pairs already present in temperatures higher than $T_c$  ~\cite{anzai}. 
In that case, the $T^*(p)$-line should follow closely the superconducting dome in the overdoped region, vanishing together with superconductivity 
when the $T_c(p)$-line reaches zero ~\cite{kordyuk, hashimoto}.
A second scenario leads to interpretations where the pseudogap is attributed to excitations different from those related to 
superconductivity ~\cite{tallon-loram}; in this case the $T^*(p)$-line should cut the superconducting dome ending in a quantum critical point (QCP)
at some critical value $p_{cp}$, smaller than that defining the upper limit of the dome.
A third and more recent scenario is also based on the competition between a charge or spin ordering phenomenon and superconductivity,
but without the need of a QCP. These are the charge density wave, spin density wave ~\cite{tigran, charlebois} and spin-charge
separation ~\cite{chingkit, taikai} scenarios.

All the above scenarios point to the same characteristics of the $T^*(p)$-line in the normal phase of underdoped specimens. 
However, divergences arise when dealing with the superconducting phase and even with the normal phase of both under and overdoped samples  ~\cite{hashimoto}.
Regarding the recently found experimental characteristics of the pseudogap in the normal phase, high precision torque-magnetometer experiments  ~\cite{sato} 
and optical measurements ~\cite{zhao} on YBa$_2$Cu$_3$O$_x$ (YBCO) samples detected evidences for broken rotational (nematic) and inversion
(odd-parity magnetic phase) symmetry phases below $T^*$, respectively. 
Both of these results are in favor of the QCP scenario. Additionally, muon spin rotation measurements in YBCO led to the observation of slow 
magnetic fluctuations at temperatures close to $T^*$, indicating that the pseudogap is an authentic thermodynamic phase stabilized
due to intra-cell spin ordering  ~\cite{zhang}. For the Bi-2212 system, a recent report using 
ultrahigh resolution resonant inelastic X-ray scattering showed evidence of dispersive charge density waves 
(CDW) closely related the pseudogap ~\cite{chaix}. Notwithstanding the advances made to detect and understand 
the trends of the pseudogap in the normal phase of the HTSC, the effects and consequences of this property inside the superconducting phase remain 
scarcely explored. Recent attempts to prove the interplay between the charge ordering and pseudogap 
effects in temperatures below $T_c$ include the destruction of the superconducting state by using high magnetic fields 
~\cite{badoux, sebastian} and the employ of sophisticated techniques, as ultrasound spectroscopy ~\cite{shekhter}.

Here we report on possible evidences of pseudogap effects in the superconducting phase of
Bi$_2$Sr$_2$CaCu$_2$O$_{8+\delta}$  (Bi-2212) single crystals with different oxygen concentrations. 
We do this by analyzing the kinetic energy density ($E_k$) that is induced in the superconducting charge carriers of a type-II superconductor when it is submitted to a magnetic field in the vortex region of the phase diagram. Experimentally, this  field-induced kinetic energy density may be obtained from the product of the 
equilibrium magnetization ($\mathbf{M}$) and the magnetic induction ($\mathbf{B}$) in the reversible region
where no pinning effects are present. Mathematically, one writes  ~\cite{doria}:
\begin{equation}\label{eq_Ek_calculation}
 E_k=-\mathbf{M}\cdot\mathbf{B}.
\end{equation}

Equation (\ref{eq_Ek_calculation}) is a consequence of the classical idea of the virial theorem applied to the Ginzburg-Landau (GL) theory ~\cite{doria-gubernantis}. Specifically, an spacial scale transformation is applied to the GL's
free-energy where both the 
vector potential and the order parameter are transformed. By implementing boundary conditions for these two parameters,
and by minimizing the energy with respect to the scale parameter, the expression 
$\mathbf{H}\cdot\mathbf{B}=4\pi(E_k+2E_f)$ (where $E_k$ and $E_f$ are the kinetic and magnetic-field energy densities, respectively) is obtained. This expression is the so called 
``Virial Theorem of Superconductivity'' \cite{doria-gubernantis} . 
Finally, by expressing the magnetic induction in terms of $\mathbf{H}$ and $\mathbf{M}$, and by utilizing
a mean-field approximation for $\mathbf{H}$, the virial theorem can be re-arranged to yield $E_k$ as seen in Eq. (\ref{eq_Ek_calculation}) ~\cite{doria}. 
This result represents the excess of kinetic energy density induced by an the external magnetic-field upon the superconducting pairs of an anisotropic Type-II superconductor in the equilibrium regime ~\cite{doria, doria-sugui-oliveira}.

The study of $E_k$ has been useful to reveal the vortex dynamics of the superconducting condensate 
in different high temperature superconducting cuprates (HTSC) ~\cite{doria-sugui-oliveira, sugui, penaSm, vieira}.
Here we show how the analysis of $E_k$ can also let us to indirectly investigate the nuances of the DOS at Fermi level and thus to obtain information on the pseudogap phenomenon from the inside of the superconducting phase.
From this analysis we find that, in the underdoped regime, the amplitude of $E_k$ at fixed field and reduced temperature $T/T_c$ decreases steadily with decreasing holes density, consistently with expectations for a decreasing DOS at the Fermi level in that regime. 
Going further, $E_k$ goes through a sharp maximum at $p \sim 0.18$, in the slightly overdoped regime.
We ascribe this peak to a van Hove singularity related to the sudden closing of the pseudogap phase occurring at this carriers density.
This result is in accordance with the scenario where the pseudogap is related to excitations different from those giving origin to the superconducting state.  
Finally, fittings of the ratio $E_k/(\mu_0H)$ to $(\mu_0H)$ let us to obtain the superfluid density $\rho_s$.
When $\rho_s$ is plotted as a function of the hole concentration, it reveals a deviation of the 
Uemura relation around the optimal doping. 
Consequently, the values of $p$ where the critical temperature and the superfluid density are maximal do not coincide.
We attribute the deviation of the Uemura’s law to the presence of pseudogap phase. 

\section{Materials and Methods}

Several Bi$_2$Sr$_2$CaCu$_2$O$_{8+\delta}$ single crystals were synthesized by the the auto-flux method  following the 
procedure described in Ref. ~\cite{Bi2212_prep_Luti, Bi2212_prep_Luti_MatToday}.
The crystals have the form of platelets with size between 1 - 3 mm and thickness around 30 $\mu$m. 
The hole density of the as-grown crystals was modified by thermal treatments at fixed temperatures 
in vacuum or oxygen atmosphere to remove or add oxygen, respectively ~\cite{Bi2212_prep_Luti, Bi2212_prep_Luti_MatToday}.
The quality and uniqueness of the crystallographic phase was probed, for each sample, by X-rays diffraction measurements 
performed with a  SIEMENS D5000 diffractometer equipped with a cooper anode.
The obtained diffraction patterns showed only even and sharp (00l) peaks.

\begin{figure}
\centering
 \includegraphics[keepaspectratio,width =7truecm]{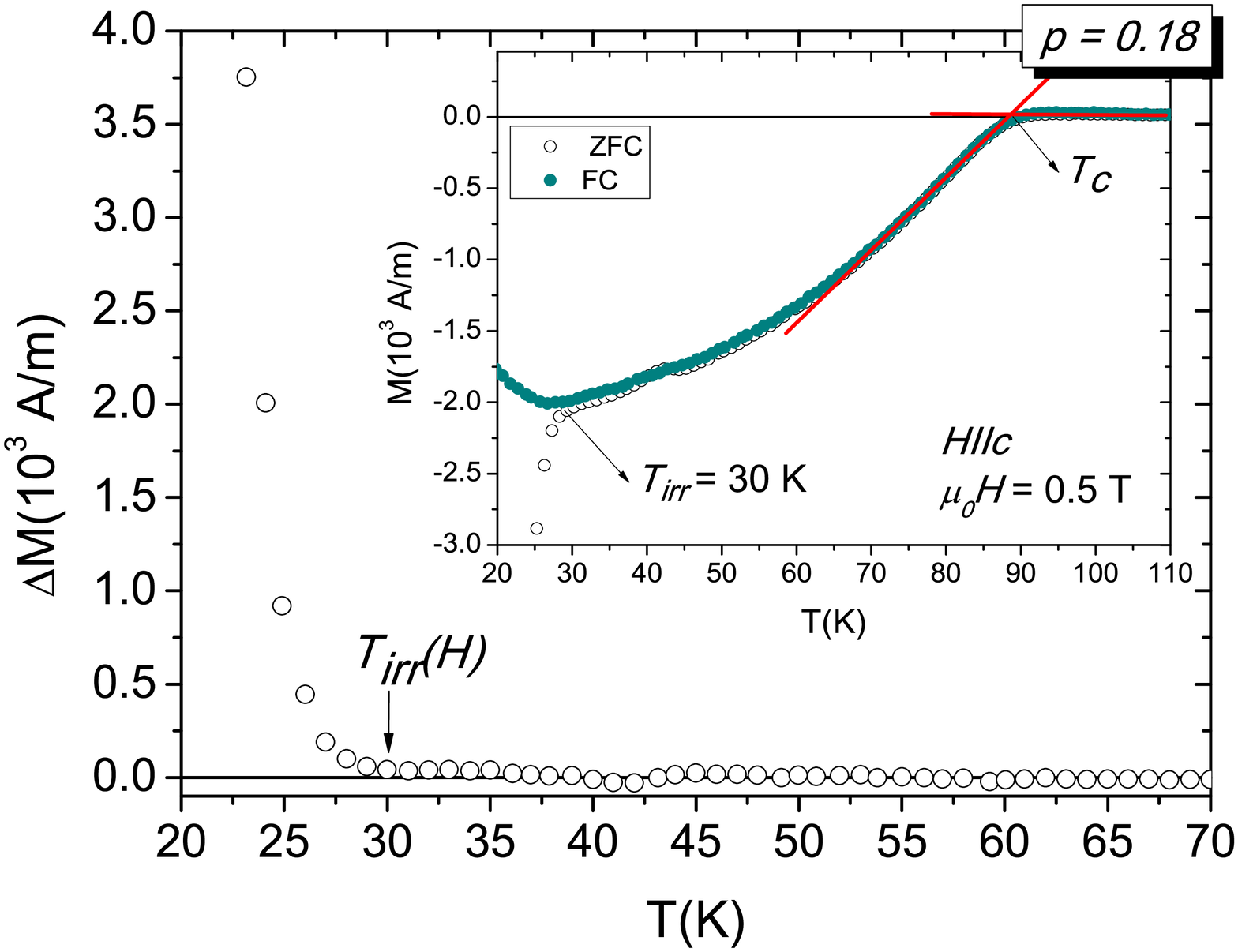}
 \caption{Representative curve of the difference between the zero field cooled (ZFC) and field cooled (FC) magnetic moments ($\Delta M= M_{FC}-M_{ZFC}$)
 for a Bi-2212 sample in the optimally doped regime. $T_{irr}$ is the temperature where this difference becomes larger than zero.
 The inset shows graphically the criterion used for determining $T_c$ at each value of applied magnetic field. There
 the $T_{irr}$ is also shown as the temperature  where the ZFC and FC curves split apart.}\label{Tc_Tirr}
\end{figure}

Zero field cooling (ZFC) and field cooling (FC) magnetization measurements were carried out in fields
ranging from $\mu_0H=1$ mT up to $\mu_0H=500$ mT  in the configuration where the field is applied parallel to the $c-$axis.
A  Quantum Design XL5-MPMS SQUID magnetometer was used in these experiments.
The critical temperatures reported here for all samples were extracted from the  ZFC curves at $\mu_0H=1$ mT; $T_c(H)$
was estimated as the intersection point of two straight lines fitted to the data in the normal
and superconducting phases; this procedure is exemplified in the inset of Fig. \ref{Tc_Tirr} for a different applied field 
($\mu_0H=0.5$ T). 
The irreversibility temperature $T_{irr}$ was also obtained from each pair of ZFC and FC magnetizations. 
This  characteristic temperature  is the lowest limit of the temperature range where the equilibrium magnetization
is straightforwardly obtained from the data.
The irreversibility temperature for each value of magnetic field was defined by the point where the difference between the 
FC and ZFC magnetizations becomes immeasurably small. 
The locus of $T_{irr}$ for a fixed magnetic field $\mu_0H=500$ mT is exemplified in the main frame of Fig. \ref{Tc_Tirr}.

The experimental magnetization between $T_{irr}$ and $T_c$ was used for obtaining the kinetic energy density
with the help of Eq. \ref{eq_Ek_calculation}, as will be shown in the next section.
The corresponding  magnetic induction was calculated as
\begin{equation}
 B=\mu_0[H_{ap}-(1-\eta) M],
\end{equation}
 where  $\mu_0$ is the vacuum permeability,
$H_{ap}$ is the applied field in A/m, $M$ is the magnetization and $\eta$ is the geometric demagnetization factor. 
This factor was estimated with basis on the calculations presented in Ref. ~\cite{osborn} 
and by approximating the samples' shape by ellipsoids.
The hole concentrations $p$ were calculated using the quadratic, empirical relation ~\cite{tallon-bernhard}:
\begin{equation}\label{eq_Tc_calc}
  T_c = T_{c, max}[1 - 82.6(p - 0.16)^2].
 \end{equation}
 
\begin{figure}
\centering
 \includegraphics[keepaspectratio,width =7truecm]{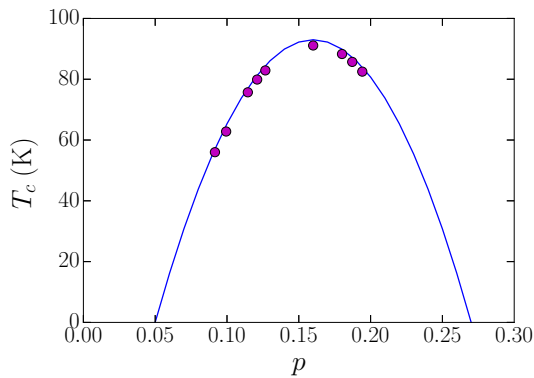}
 \caption{Superconducting critical temperature obtained from magnetization curves measured at $\mu_0H=1$ mT as a function of the hole doping for all the samples studied  in this work (points). Error bars are
 within the size of the symbols.  The continuous line is a fitting to Eq. (\ref{eq_Tc_calc}).}\label{domo_exp}
\end{figure}

The superconducting dome as obtained from the $T_c(p)$ data for our samples and from fittings to Eq. (\ref{eq_Tc_calc})
is shown in Fig. \ref{domo_exp}. 

\section{Results and discussion}

\begin{figure}[h!]
 \centering
 \includegraphics[keepaspectratio,width =8truecm]{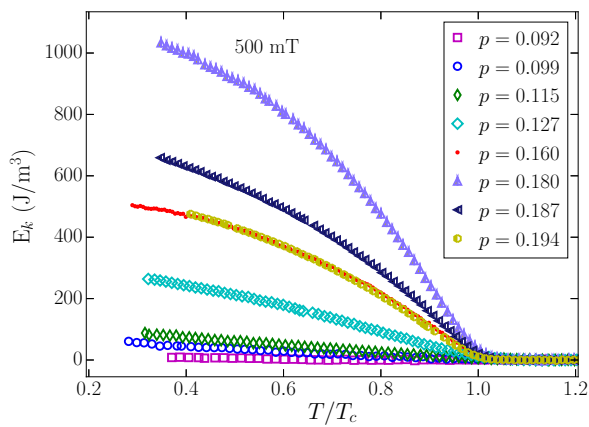}
 \caption{\small{Kinetic energy density in a constant field $\mu_0H = 500$ mT as a function of the normalized temperature 
 for Bi-2212 crystals with the quoted carrier concentrations.  
 The highest kinetic energy density is observed for the sample with $p=0.180$ (see Fig. \ref{Ek_vs_p})}}\label{Ek_vs_T}
\end{figure}

Figure \ref{Ek_vs_T} shows the calculated $E_k$ as a function of the reduced temperature ($T/T_c$) for a constant magnetic-field 
\mbox{$\mu_0H=500$ mT}. 
One observes that $E_k$ extrapolates to zero at the critical temperature in all cases.
At most, weak rounding effects due to thermal fluctuations are perceived in the close vicinity of $T_c$.
Previous studies in optimally and underdoped samples of YBCO
(YBCO) ~\cite{sugui}, optimally doped Bi-2212 ~\cite{sugui} and La$_{1.9}$Sr$_{0.1}$CuO$_4$ ~\cite{doria-sugui},
reported that an appreciable amount of $E_k$ subsists above $T_c$. 
The kinetic energy excess found by  authors in Refs. ~\cite{sugui} and ~\cite{doria-sugui} was interpreted by them
as resulting from non-correlated Cooper pairs characterizing the pseudogap phase. 
Results in Fig. \ref{Ek_vs_T}, however, are rather indicating that the dependence of  $E_k$  with field and temperature is 
mostly related to the superconducting gap, as expected with basis on the BCS theory. We note that the observation of a 
certain  $E_k$ amplitude above $T_c$ can be alternatively explained as an effect of strong thermal fluctuations.

\begin{figure}[h!]
\centering
 \includegraphics[keepaspectratio, width= 8.0truecm]{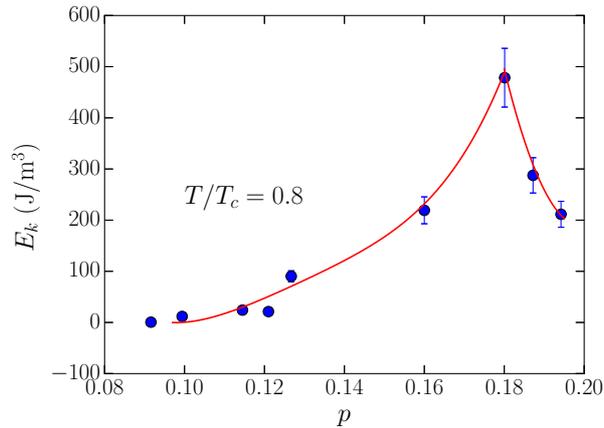}
 \caption{\small{Amplitude of the kinetic energy density at the reduced temperature $T/T_c=0.8$ as a function of the holes concentration. 
 The continuous line is a guide for the eyes.}}\label{Ek_vs_p}
\end{figure}

We plot in Fig. \ref{Ek_vs_p} the amplitude of $E_k$  measured for each sample at $T = 0.8 T_c$  and $\mu_0H=500$ mT as a function of the holes density $p$.
A prominent feature in Fig. \ref{Ek_vs_p} is the sharp maximum of $E_k$ that occurs at $p = 0.18$. This value for $p$ 
does not match with the one where the maximal critical temperature is observed ($p = 0.16$). Moreover, the $E_k$ vs. $p$ curve does
not follow the same dome-like shape of the $T_c$ vs. $p$ curve, as expected from the Uemura relation. Then, the observed peak in $E_k$ suggests 
that some unusual variation of the properties ruling the kinetic energy density in the superconducting state of Bi-2212 must be taken into account.

The superconducting gap, identified in Angle Resolved Photo-emission Spectroscopy (ARPES) around the nodal region of the Brillouin
zone of Bi-2212, is nearly independent of the carrier concentration in the underdoped region of the 
dome-like phase diagram  ~\cite{hashimoto, tallon-loram,  tanaka}. Assuming the validity of this observation, we infer that the
dependence of the measured kinetic energy density with the carriers concentration shown in Fig. \ref{Ek_vs_p} does
not follow the superconducting gap and might be at least partially attributed to some different phenomenon. 
On the other hand, the results in Fig. \ref{Ek_vs_T} suggest that the temperature and magnetic-field dependences of $E_k$ are mainly 
ruled by the superconducting order parameter, as expected. Thus, the behavior of $E_k(H, T, p)$  in our Bi-2212 samples
suggests that some distinct electronic phase coexists with the superconducting state below $T_c$.

Motivated by the possibility of obtaining further insight on the validity of the above outlined interpretation,
we analyze in detail the behavior of $E_k$ as a function of the magnetic field. 
As in Ref. ~\cite{penaSm}, 
we assume that the magnetization of our samples is well described by 
the London approximation to the GL theory; this approximation is valid in the low field region 
where vortices do not overlap significantly. Then, the London equation for $\mathbf{M}$ was used to obtain   
$\mathbf{B}$, and both were replaced in Eq. (\ref{eq_Ek_calculation})  which remains valid within this context  ~\cite{doria, doria-sugui-oliveira}.
We deduce the following expression for $E_k$ ~\cite{penaSm}: 
\begin{equation}\label{eq_Eknor_H}
  \frac{E_{K}(\mu_0H)}{\mu_0H}=\frac{\phi_0}{8\pi\lambda^2\mu_0}\ln\frac{\beta_L\mu_0H_{c2}}{\mu_0H}-
\left(\frac{\phi_0}{8\pi\lambda^2}\right)^2
\frac{1}{\mu_0^2H}\left(\ln\frac{\beta_L\mu_0H_{c2}}{\mu_0H}\right)^2,
\end{equation}                               
where $\phi_0$ is the  magnetic flux quantum, $\mu_0$ is the vacuum permeability, $\lambda$ is the London penetration depth,
$H$ is the applied field, $H_{c2}$ is the upper critical field, and $\beta_L$ is a geometrical parameter of order unity.

\begin{figure}
\centering
 \includegraphics[keepaspectratio, width= 7.8truecm]{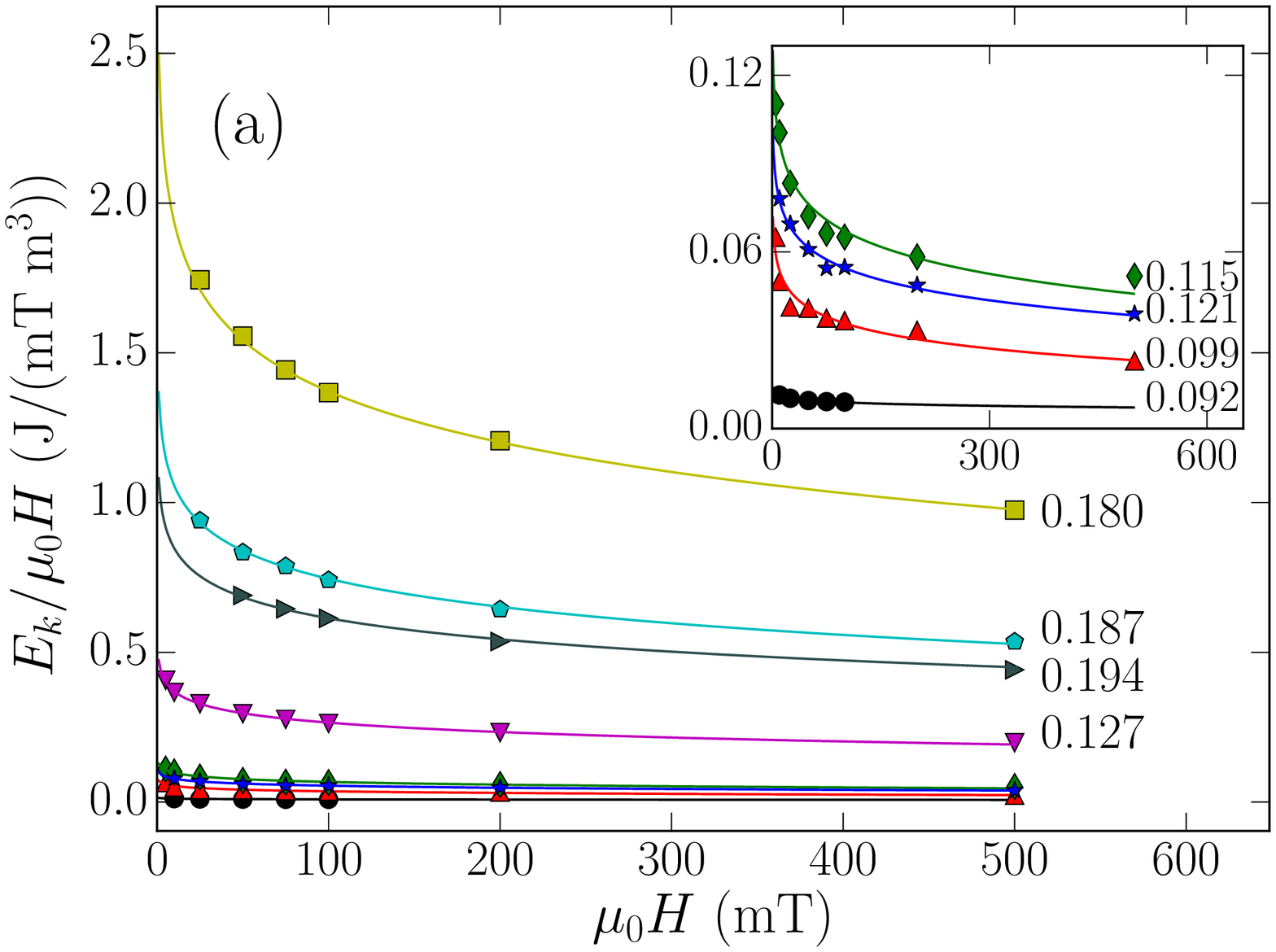}
 \includegraphics[keepaspectratio, width= 8.0truecm]{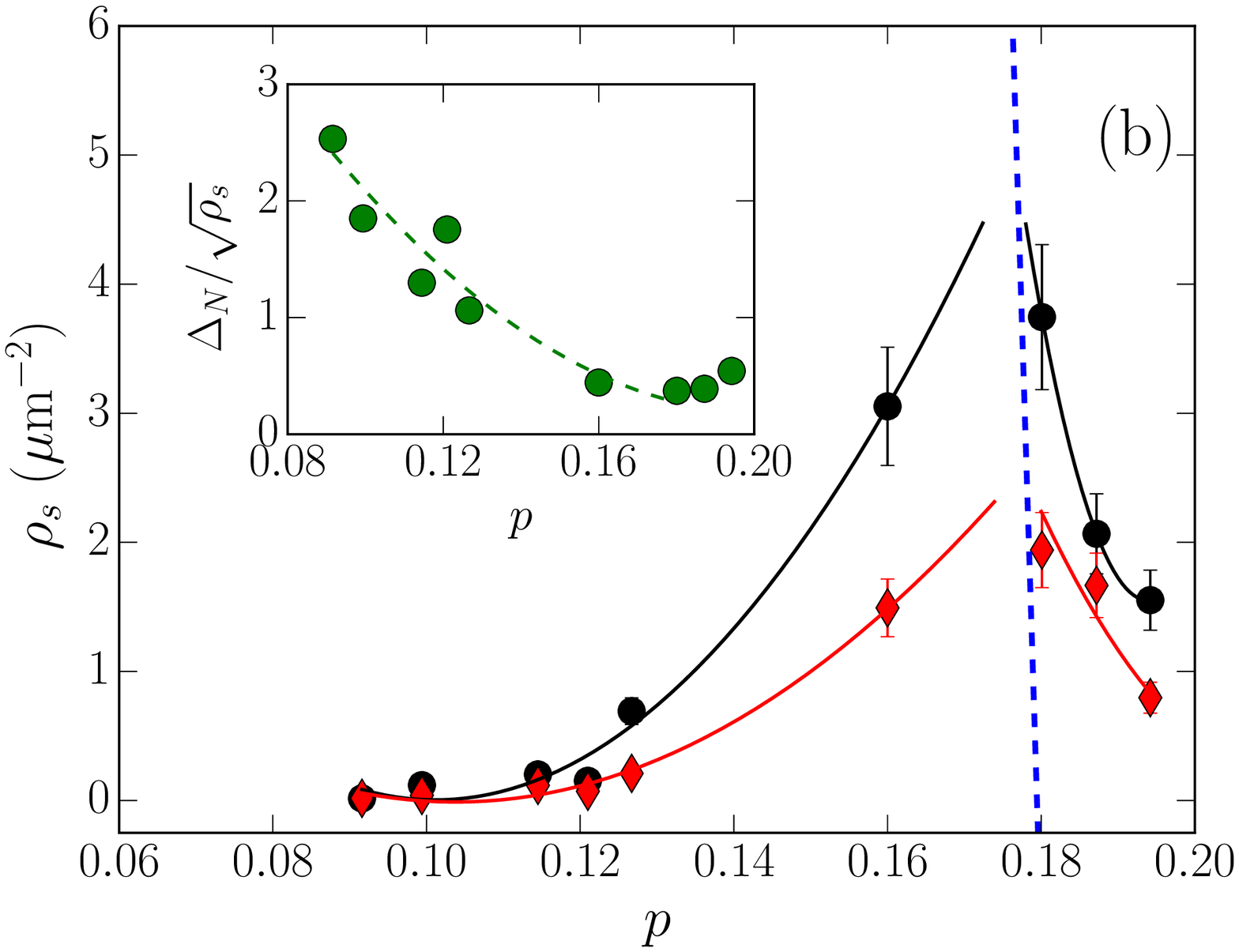}
 \caption{\small{(a) $E_k/(\mu_0H)$ vs. $\mu_0H$ for most of the studied samples ($p$ values are indicated for each curve);
 symbols represent the experimental points and  continuous lines are fittings to Eq. \ref{eq_Eknor_H}.
 (b) Superfluid density $\rho_s = (1 / \lambda^2)$ as a function
 of the holes concentration for  $T / T_c = 0.8$ (rounded symbols) and $T / T_c = 0.9$ (diamond symbols).
 The continuous lines are  guides for the eyes. The straight dashed line shows that the maximum of the $\rho_s$ vs. $p$ curves extrapolates to $p\sim0.18$.
 In the inset of panel (b) the ratio between the superconductig gap and the square root of the superfluid density as 
 calculated from Eq.~\ref{deltax} is represented as a function of $p$ for $T / T_c = 0.9$ (see text). The dashed line is a guide for the eyes. 
}}\label{sdensity_vs_p}
\end{figure}

Fittings of the experimental points to Eq.  (\ref{eq_Eknor_H}) were performed by using 
a PYTHON  program to extract both the higher critical field and the penetration depth for each sample at some fixed temperatures.
The fitting results are presented as solid lines in Fig. \ref{sdensity_vs_p}(a) which is for a reduced
temperature $T / T_c = 0.8$ for all samples. The fitting parameters for this case are displayed in Table \ref{fitting_par}.

\begin{table}[h]
\caption{Fitting parameters of the experimental points to Eq. (\ref{eq_Eknor_H}) for all
samples at $T / T_c = 0.8$.}\label{fitting_par}
\begin{center}

\begin{tabular}{cccc}\hline
\textsl{p} ($\pm 0.001$) & $T_c$ ($\pm 0.5$) K &  $\lambda$ ($\pm 0.05$) $\mu$m & $H_{c2}$ ($\pm 0.5$) mT \\
\hline
0.092	&	56.0	&	7.85	&	11.7\\
0.099	&	62.8	&	2.90	&	9.8\\
0.115	&	75.7	&	2.22	&	15.8\\
0.121	&	79.9	&	2.56	&	23.4\\
0.127 	&	82.9	&	1.20	&	34.8\\
0.160   &   92.3	&   0.57   &    0.54\\
0.180	&	88.3	&	0.52	&	22.3\\
0.187	&	85.7	&	0.69	&	24.8\\
0.194	&	82.5	&	0.80	&	41.9\\
\hline 
\end{tabular}
 
\end{center}

\end{table}

These extracted  values for $\lambda$ were used to calculate the superfluid density from ~\cite{tallon-loram-cooper}:
\begin{equation}\label{rho_s}
\rho_s=\frac{m}{2\mu_0e^2}\frac{1}{\lambda^2}.
\end{equation}

We plot in the main panel of Fig. \ref{sdensity_vs_p}(b) the superfluid density
as a function of the hole concentration for the fixed reduced temperatures $T / T_c = 0.8$ (rounded symbols) and $T / T_c = 0.9$ (diamond symbols). 
As we are mostly  interested in the overall qualitative behavior than in the numerical values of $\rho_s$, we assume $m/(2\mu_0e^2)=1$.
The result in Fig. \ref{sdensity_vs_p}(b) is very similar  to that found from ARPES measurements in Ref. ~\cite{storey2008}
and basically reproduces the behavior of $E_k$ as a function of $p$ shown in Fig. \ref{Ek_vs_p}. Though interesting, 
the similitude between results in Figs. \ref{Ek_vs_p} and \ref{sdensity_vs_p}(b) is not really surprising within the GL theory context.
There, the relation  $E_k \propto \langle\arrowvert\psi\arrowvert^2\rangle$, where $\psi$ is the superconducting order parameter
 and $\langle ... \rangle$ symbolizes a spatial average ~\cite{sugui}, is satisfied. Consequently,   the proportionality between
 the density of superconducting pairs, $\arrowvert\psi\arrowvert^2$,  and $\lambda^{-2}$ is expected.
Finally, the comparison of  Figs. \ref{domo_exp} and \ref{sdensity_vs_p}(b) lets one to perceive  that, as mentioned in the introduction,
the series of Bi-2212 single crystal samples studied here violates the phenomenological Uemura's law which foresees a linear relation between 
$\rho_s$ and  $T_c$. In the underdoped regime, the graph $\rho_s$ vs. $T_c$ has a sub-linear behavior (not shown), qualitatively
similar to that of the $\rho_s$ vs. $p$ in Fig. \ref{sdensity_vs_p}(b). The violation of the Uemura’s law nearby the optimal doping had already been reported by C. C. Homes \textsl{et. al.} in Ref. ~\cite{homes},
where the proportionality $\rho_s\propto\sigma_{dc}T_c$ ($\sigma_{dc}$ is the conductivity measured near $T_c$)
was proposed as the appropriate scaling to substitute the Uemura relation in all families of superconducting cuprates. 
On the other hand, as a further exemple of violation of the Uemura relation, J. Hetel \textsl{et. al.} ~\cite{hetel} 
found a sublinear relation between $T_c$ and the superfluid density in the strongly underdoped region of thick YBCO films.

The superfluid density $\rho_s$ may be conceived as a measure of the  toughness of the superconducting state against an external magnetic field.
As the condensation energy is proportional to the critical field, tougher superconducting states have higher condensation energies
and superfluid densities.
In this sense, it is remarkable that the maximum of  the $\rho_s$ vs. $p$ curves extrapolates to $p\sim0.18$, as indicated by
the straight, dashed line in the main panel of Fig. \ref{sdensity_vs_p} (b)). 
This is consistent with results of thermodynamic measurements that show a steep increase of the condensation energy  at $p\sim0.19$ 
in Y$_{0.8}$Ca$_{0.2}$Ba$_2$Cu$_3$O$_{7-\delta}$ ~\cite{tallon-loram}.
Additionally, the direct proportionality between $E_k$ and $\rho_s$ lead us to straightforwardly conclude that the doping dependence
of both of these quantities indicates that a sharp increase in the density of superconducting carriers occurs at  $p=0.18$.
This conclusion agrees with the fact that, this same maximum is also present in the DOS
of other HTSC systems, as estimated in Ref. ~\cite{storey} with basis on experimental results.
The qualitative coincidences in the behavior of the kinetic energy density, entropy ~\cite{storey2008}, 
superfluid density \cite{tallon-loram-cooper}, and DOS ~\cite{storey}
with respect to $p$ in the Bi-2212 and other HTSC is indeed remarkable. 
All these results are compatible with the existence of a van Hove singularity (vHs) in the slightly overdoped region of the Bi-2212 phase diagram. 
Such singularity has been related to the steep ending of the pseudogap phase ~\cite{hashimoto, storey2008} occurring simultaneously with a
Lifshitz quantum phase transition where the active hole-like Fermi surface becomes electron-like ~\cite{fujita, benhabib}.
This is accompanied by a notable transition in $k-$space topology  within the narrow range $p\sim0.19\pm0.1$  ~\cite{fujita}.
Within this frame, the peak observed at $p\sim0.18$ in our \mbox{$E_k$ vs. $p$} and $\rho_s$ vs. $p$ curves 
suggests  that, differently from $T_c$, the kinetic energy density and the superfluid density
are rather related to the pseudogap energy scale.

It was found experimentally that in Bi-2212 the pseudogap energy ($\Delta^\ast$) satisfies the relation ~\cite{anzai}:
\begin{equation}\label{deltax}
 \Delta^\ast\propto\frac{\Delta_N}{\sqrt{\rho_s}},
\end{equation}
where $\Delta_N=4.25k_BT_c$  ~\cite{anzai} is the superconducting (nodal) gap. The quotient $\Delta_N/\sqrt{\rho_s}$ 
estimated from our data  is presented as a function of $p$ in 
the inset of Fig. \ref{sdensity_vs_p} (b).
Assuming the validity of Eq. (\ref{deltax}),  we interpret the minimum of $\Delta_N/\sqrt{\rho_s}$ observed at $p=0.18$ as a consequence of a minimum 
in the pseudogap energy. 
This interpretation is consistent with our previous statement on the existence of a maximum in the density of superconducting carriers at the same $p$ value.
As mentioned before, that maximum is consequence of a Lifshitz transition apparently driven by the reduction of the strength of the electronic 
correlations with doping ~\cite{braganca}.
Thus, our results indicate that the reinforcement of the superfluid and kinetic energy densities in the superconducting state
is related to the undermining of the excitations that give origin to the pseudogap phenomenon.

Our kinetic energy density results indirectly support the existence
of a QCP by $p = 0.18$, in the lightly overdoped region of the Bi-2212 phase diagram;
even so, these results don't rule out the possibility of the occurrence of the pseudogap in samples
with $p\gtrsim0.19$ in temperatures outside the superconducting dome. 
In fact, effects of the pseudogap up to $p\sim0.22$ were observed in ARPES ~\cite{hashimoto, hashimoto2} and high-field NMR ~\cite{kawasaki}  measurements performed on samples of the Bi$_2$Sr$_2$CuO (Bi-2201) and Bi-2212 systems. 
In Refs.  ~\cite{hashimoto} and ~\cite{hashimoto2} the effects of pseudogap were observed 
up to $p\sim0.22$ in temperatures above $T_c$; there, the authors put forward 
a phase diagram where the pseudogap goes into the superconducting dome drawing a positive-slope line starting at
$p\sim0.22$ and ending in a QCP at $p\sim0.19$ over the horizontal axis. 
In Ref. ~\cite{kawasaki} ultra-high magnetic fields were used to suppress the superconductivity to barely observe the pseudogap phenomenon. 
Our results show that the temperature-dependence of $E_k$ is dominated by the superconducting order parameter, while its doping-dependence is rather dominated by the pseudogap.
All these scenarios are consistent with a picture where the superconducting and pseudogap phases coexist,
but they compete at the point that for samples with $p>0.19$ the pseudogap is no longer able to manifest itself
overwhelmingly inside the superconducting dome. 
To finish this point, the fact that our results support the location of the QCP nearer, and even a little lower
than $p\sim0.19$ may be due to the fact that we are approaching the  pseudogap from the point of view of
an intrinsically superconducting parameter in which the pseudogap is already manifesting more weakly.

Another possible evidence of  competition among different  electronic states in Bi-2212  comes from a 
study of the normal-phase susceptibility in the same set of samples investigated in the present work ~\cite{Bi2212_prep_Luti}.
Those results showed that a maximum in the DOS occurs at $p\sim0.16$ and not at $p\sim0.18$.
The discrepancy between these two characteristic values for $p$ is probably consequence  of the used experimental techniques.
Depending on the temperature range where these methods are implemented, some are more sensitive to the effects related to the nodal
region of the Brillouin zone, others to the antinodal region.

  
%

\section{Conclusion}

We studied the overall behavior of the in-field kinetic energy density in a series of Bi-2212 single crystals with different carrier density.
Results of $E_k$ as a  function of the temperature, doping and magnetic field were analyzed  with 
the aim of identifying features related to the pseudogap phenomenon. 
From the $E_k$ vs. $T$ curves, we conclude that the field-induced kinetic energy changes with temperature similarly to
the superconducting order parameter.
At fixed temperature and variable fields, $E_k$ is quite well described by the London approximation to the GL theory.
On the other hand, the variation of $E_k$ with doping can not be explained solely with basis on the superconducting order parameter.
In samples with  $p < 0.18$, both the kinetic energy  and superfluid densities are strongly depressed.
This behavior is expected in properties which are closely dependent on the pseudogap.
A sharp maximum is observed  in  $E_k$ at the carrier concentration $p = 0.18$ (also evident in the superfluid density). 
This particular feature is consistent with  with the occurrence of a van Hove singularity in the DOS coincident with the suppression of the pseudogap.

The fact that $E_k$ is ruled by both the superconducting and pseudogap energy scales strongly suggests that the 
superconducting state and the pseudogap phenomenon coexist inside the superconducting dome for all Bi-2212 samples with $p\leq0.18 $.
This general conclusion is consistent with a model that attributes essentially different origins
for the superconducting state and the pseudogap phenomenon.
A comparison between our results and other experimental data with theoretical analyses lead us to conclude that
the pseudogap phenomenon influences the behavior of some quantities intrinsically related to superconductivity in the HTSC.
Consequently, one may expect that its effects also occur  inside the superconducting dome. 
%


\section*{Acknowledgments}

This work was partially financed by the Brazilian agencies FAPERGS and CNPq (Grant PRONEX 16/0490-0). 
L. F. Lopes benefits from a CNPq fellowship.
J. P. Pe\~na received a post-doctoral fellowship of the Brazilian Agency CAPES.

\end{document}